\documentclass[apj]{emulateapj}\usepackage{apjfonts}

\shorttitle{AGB stars in MSX and IRAS colors} 
\shortauthors{Sjouwerman, Capen \& Claussen}

\begin{document}

\title{MSX versus IRAS Two-Color Diagrams and the 
CSE-Sequence of Oxygen-Rich Late-Type Stars}

%

\author{Lor\'ant~O.~Sjouwerman\altaffilmark{1},
Stephanie~M.~Capen\altaffilmark{1, 2},
Mark~J.~Claussen\altaffilmark{1}}

\altaffiltext{1}{National Radio Astronomy Observatory, P.O.~Box O,
Socorro, NM 87801}
\altaffiltext{2}{Eastern Nazarene College, 23 East Elm Avenue,
Quincy, MA 02170}

\email{lsjouwerman@nrao.edu, stephanie.m.capen@enc.edu, mclaussen@nrao.edu}

\begin{abstract}

We present MSX two-color diagrams that can be used to characterize
circumstellar environments of sources with good quality MSX colors in
terms of IRAS color regions for oxygen-rich stars. 
With these diagrams we aim to provide a
new tool that can be used to study circumstellar environments and to
improve detection rates for targeted surveys for circumstellar maser
emission similar to the IRAS two-color diagram. This new tool is
especially useful for regions in the sky where IRAS was confused, in
particular in the Galactic plane and bulge region. Unfortunately,
using MSX colors alone does not allow to distinguish between
carbon-rich and oxygen-rich objects. An application of this tool on 86
GHz SiO masers shows that for this type of masers an instantaneous
detection rate of 60\% to 80\% can be achieved if target sources are
selected according to MSX color (region).

Our investigations may have revealed an error in the MSX point source
catalog version 2.3. That is, the photometry of the 21.3 $\mu$m (MSX E
filter) band for most weak 8.28 $\mu$m (or MSX A filter) band sources
seems off by about a factor two (0.5--1 magnitude too bright).

\end{abstract}


\section{Introduction}\label{intro}

The two-color diagram (2CD) of stellar dust/gas envelope sources in
the Infrared Astronomical Satellite (IRAS) point source catalog (PSC)
has been very useful in characterizing the circumstellar environment
of late-type stars, typically thought to be at the top of the
Asymptotic Giant Branch (AGB) \citep{neugebauer84, beichman85,olnon84,
hacking85,vdveen88}. The IRAS 60$\mu$m/25$\mu$m versus
25$\mu$m/12$\mu$m two-color diagram statistically distinguishes between
carbon-rich and oxygen-rich objects. In the latter color it also indicates
the thickness of the circumstellar shell based on the appearance of the
9.7 $\mu$m silicate feature in the individual 7.7 to 22.6 $\mu$m
low-resolution spectra (LRS) \citep{olnon86,vdveen88}.

For many studies since the IRAS mission, the IRAS 2CD has thus been the 
starting point for color-selection of samples; not only to select
AGB or Red Giant Branch (RGB) stars, but also to select e.g.\ proto-planetary
nebulae, young stellar objects and starburst galaxies. The IRAS 2CD
has been revolutionary in the study of the late evolutionary stages of
evolved stars and chemical enrichment of the Galaxy due to the
enhanced mass-loss of these stars. Moreover, the 2CD selection is very effective in 
finding circumstellar masers, and hence line-of-sight velocities of
stars that then can be used for kinematic studies of the Galaxy. For
example, objects in IRAS color region IIIb and IV \citep{vdveen88} have been
targeted for 1612 MHz OH maser emission, whereas objects in region II usually have
a circumstellar envelope (CSE) too thin to form and sustain these OH
masers. On the other hand, objects in region II may be CSEs in which SiO maser
emission is more likely to exist \citep[e.g.,][and the extensive list
of references therein]{habing96}.

The IRAS all-sky survey has a significant shortcoming in the sense that due
to the low angular resolution, the Galactic plane (and in
particular the bulge) survey region suffers from confusion. This is
unfortunate as the majority of evolved stars (but also e.g.\ young
stellar objects),  are found in the plane. Therefore subsequent infrared
surveys using the Infrared Space Observatory (ISO, with the ISOGAL
survey), the Midcourse Space Experiment (MSX, with the Galactic Plane
survey), and the Spitzer Space Telescope (with the GLIMPSE and MIPSGAL
surveys) have concentrated on the Galactic plane and bulge with much
better angular resolution\footnote{The AKARI mission is an all-sky
survey \citep{murakami07}.} \citep{omont03,price01,benjamin03,carey09}. However, 
in correcting for IRAS' shortcoming, areas of the sky which overlap with
the non-confused IRAS observations are minimal.  Also, these other surveys
have typically used other infrared wavebands and filters than IRAS.  When
the sources of interest are the intrinsically variable late-type
stars, a direct comparison of colors between IRAS and e.g.\ MSX is, 
in principle,  very difficult if not impossible.

Nevertheless, with the assumption that colors do not differ too much
between observing epochs, i.e.\ stellar phase, and thus that
results of a comparison between subsamples in a \emph{statistical}
sense can be interpreted, an attempt at a comparison between IRAS
colors and 
other infrared colors can be made. Because the PSC of
MIPSGAL and AKARI were not available in the summer of 2008 (when we began
this study), and because ISOGAL and GLIMPSE were limited in spatial coverage 
and wavelength range respectively, we here present a comparison of IRAS
and MSX colors. This comparison provides a tool to study, in the first
instance, oxygen-rich AGB and RGB stars in MSX colors, similar to the
IRAS 2CD. We note that the IRAS 2CD cannot fundamentally
be replicated in MSX colors, because the MSX data do not extend 
to a long enough wavelength so that the vertical axis of the IRAS 2CD can be 
approximated with MSX data. Most of the analysis and discussion that 
follows concentrates on the horizontal extent of the IRAS 2CD.

As an example of the capability of this comparison, we apply our
MSX color tool to investigate the occurrence of 86 GHz SiO masers in 
color-selected sources in the Galactic bulge. It is our intention to 
derive similar comparisons for Spitzer (combined GLIMPSE and MIPSGAL) and AKARI 
colors once their point source catalogs are public, and to investigate 
the occurrence of other 
CSE masers as a function of color in a similar way using these new tools.

Section~\ref{select} describes the selection of the data sets; the
sources that are in both the IRAS and MSX PSCs but excluding sources
that could confuse or contaminate a statistical analysis.  As most
masers originate from oxygen-bearing molecules, we would like to be
able to separate the carbon-rich from the oxygen-rich CSEs based
purely on MSX colors. But before investigating such a separation in
Sect.~\ref{iras2msx}, we compare the ``evolutionary curve'' of
oxygen-rich stars with circumstellar material in the IRAS 2CD as
described by \citet{vdveen88} with MSX colors. Note that this
``evolutionary curve'' is more a circumstellar envelope opacity
measure than a strict stellar mass-loss evolution curve. Therefore we
will refer to this curve as the ``CSE-sequence'' for oxygen-rich
late-type stars in the remainder of this paper. The resulting MSX 2CDs
are discussed in Sect.~\ref{msx2cd} and an example of their scientific
usefulness is shown in Sect.~\ref{sio86}.  Future plans are collected
in Sect.~\ref{plans} before closing with a summary.

\section{IRAS and MSX Data Selection}\label{select}

\subsection{Cross-identification between IRAS and MSX sources}

\begin{figure} \begin{center}
\plotone{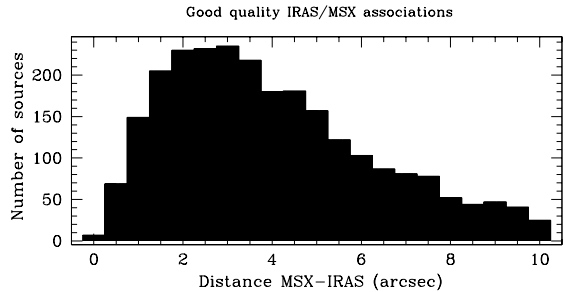}
\end{center}
\caption{Positional offsets of the 2543 cross-identifications between
  the IRAS and MSX catalogs within 10\arcsec\ radius using the
  \citet{vdveen88} IRAS 2CD boundaries and with at least one good
  quality MSX color.
\label{f1}
}
\end{figure}

For the cross-identification between IRAS and MSX sources we used the
positions from version 2.0 of the IRAS PSC (245889 sources) and Faint
Source Catalog (FSC; 173044 sources) and version 2.3 from the MSX(6C)
PSC (440487 sources) \citep{beichman88,egan03}.


Because we were very wary of contamination with incorrect
cross-identifications, but on the other hand needed a data set of
cross-identification as large as possible, we compromised between
mission position accuracy and search radius. That is, IRAS PSC 12
$\mu$m positions typically are accurate to $\sim$ 7--9\arcsec, within
a beam (angular resolution) of $\sim$ 1$\times$5\arcmin, and MSX PSC
8.28 $\mu$m positions typically are accurate to $\sim$ 2\arcsec,
within a beam of $\sim$ 18\arcsec\ \citep{neugebauer84,hacking85,
ojha07,beichman88,egan03}. We therefore used a search radius of
10\arcsec; the arithmetic mean of the beam and accuracy of the MSX mission, which
is the more restrictive between IRAS and MSX. We used the same
10\arcsec\ to cross-identify the MSX PSC with the IRAS FSC for two
reasons. First, this expands the data set with more sources to compare
IRAS and MSX colors. Second, it also reveals those IRAS-MSX PSC
cross-identifications that could be confused with faint IRAS
sources. Our data set thus excludes the MSX sources that had more than
one possible IRAS counterpart within 10\arcsec, whether or not from
the IRAS PSC or FSC, but also includes the (single) fainter IRAS
counterparts. In total we found 45895 MSX sources identified with
single IRAS PSC sources and 620 identified with single IRAS FSC sources (46515
total). Multiple identifications within 10\arcsec\ were found for 482
MSX sources, whereas 393490 MSX sources had no counterpart in the IRAS
PSC nor in the IRAS FSC within 10\arcsec\ radius.

\begin{figure} \begin{center}
\plotone{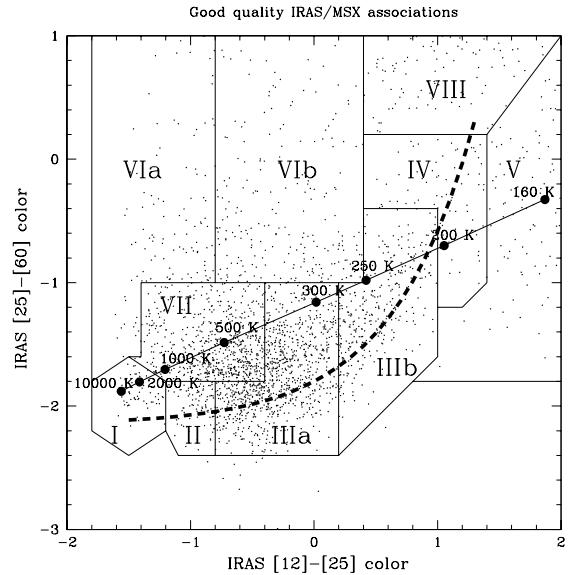}
\end{center}
\caption{The ``traditional'' IRAS 2CD with uncorrected IRAS colors,
  color regions and color ranges as in \citet[][their Fig.\
  5b]{vdveen88}. The (solid) black body line, with temperature ticks,
  and the (dashed) ``evolutionary track'' are also shown. The
  ``evolutionary track'' is referred to as the ``CSE-sequence'' in the
  remainder of this paper. The data (points) are all 2543 good quality
  IRAS sources with a likely unique MSX counterpart within 10\arcsec\
  radius for which at least one good quality MSX color can be derived.
\label{f2}
}
\end{figure}

Two MSX sources were removed because they could be (at 6.64\arcsec\ or
5.54\arcsec) identified with the same FSC source. Similarly, we
removed another 76 MSX sources for which the identification with a
unique IRAS PSC source within 10\arcsec\ was ambiguous. To be
consistent with the multiple identifications above, this means
removing the MSX source even in the case where one possible
identification was within 3\arcsec\ (2.87\arcsec) and the other was as
far as (almost) 10\arcsec\ (9.60\arcsec) away. From this sub-total of
46437 sources which match in position, we removed 26743
cross-identifications for which the MSX sources had fewer than two
reliable flux densities (i.e., ``quality numbers'' for the photometry
which were not 3 or 4), i.e., for which not a single reliable, good quality MSX
color could be calculated. Another 16194 IRAS sources were removed for
which any of the 12, 25 or 60 $\mu$m fluxes were unreliable (quality
not 3) as in that case we could not place them in the IRAS 2CD diagram
reliably. Finally, another 957 sources were removed as they fall
outside the traditional \citet{vdveen88} IRAS 2CD color magnitude
range limits (their Fig.\ 5a, 5b: $-2 < {\rm [12]-[25]} < 2$ and $-3 <
{\rm [25]-[60]} < 1$; see next subsection). The remaining catalog of
good quality IRAS sources in the traditional \citet{vdveen88} IRAS 2CD
with a likely unique MSX counterpart within 10\arcsec\ radius for
which at least one good quality MSX color could be derived is 2543.
Only two sources are from the FSC. Figure~\ref{f1} shows the histogram
of the difference in position between the final 2543 IRAS and MSX
associations. From the peaked distribution around 2.5\arcsec\ (with 
$\sim$ 18\arcsec\ angular resolution) we conclude that the
associations are very likely proper cross-identifications.  The location of
these remaining cross-identified IRAS/MSX sources in the IRAS 2CD
diagram is shown in Fig.~\ref{f2}.

\begin{figure} \begin{center}
\plotone{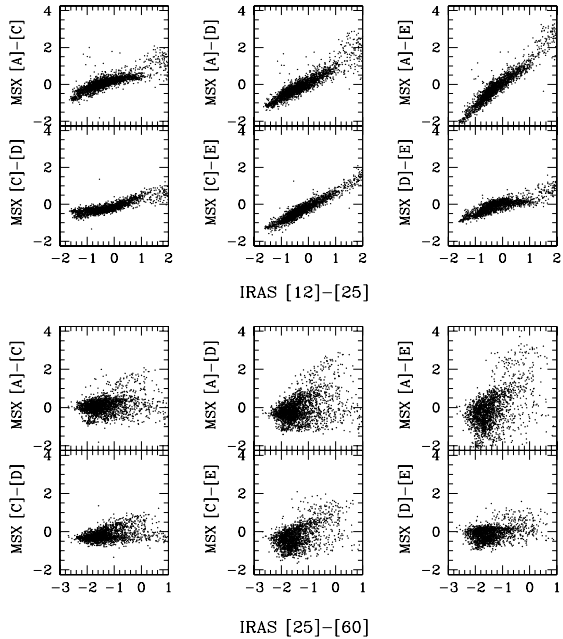}\end{center}
\caption{\small \emph{Top:} uncorrected IRAS [12]$-$[25] color versus
  combinations of the uncorrected good quality MSX colors for the 2455
  sources (2543 less 88 from IRAS region VIII) in
  Figs.~\ref{f1},~\ref{f2}. For the MSX [A]$-$[D] color, the [A]$-$[E]
  color and the [C]$-$[E] color there is a good correspondence to the
  IRAS [12]$-$[25] color, with [A]$-$[E] having a larger color range
  (a bigger stretch, steepest slope: Table~\ref{t1}). \emph{Bottom:}
  same for the IRAS [25]$-$[60] color, where no one-to-one
  correspondence between the IRAS [25]$-$[60] color can be deduced for
  any MSX color.
\label{f3}
}
\end{figure}

\begin{figure} \begin{center}
\plotone{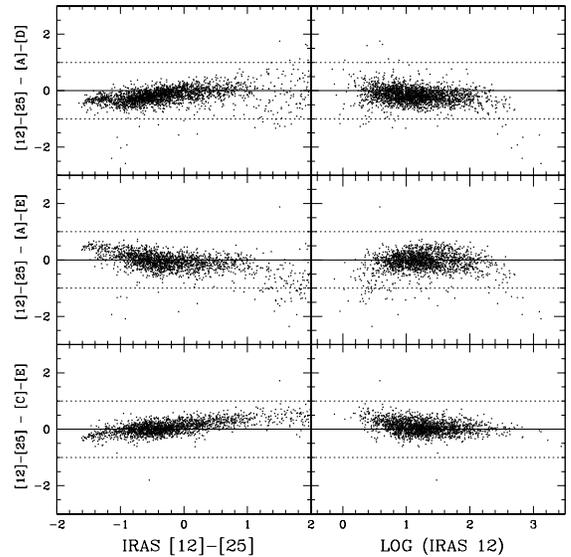}
\end{center}
\caption{IRAS [12]$-$[25] color (\emph{left}) and IRAS 12 $\mu$m flux
  (\emph{right}) versus the difference between IRAS [12]$-$[25] color
  and MSX [A]$-$[D] color (\emph{top}), MSX [A]$-$[E] color
  (\emph{middle}) and MSX [C]$-$[E] color (\emph{bottom}). The
  straight color differences are well within $\pm$1 magnitude
  (\emph{dotted lines}).  Table~\ref{t2} fits the residual dependence
  on IRAS [12]$-$[25] color.  The MSX color does not depend on total
  IRAS 12 $\mu$m flux density (neither is corrected for zero-magnitude
  offset).  Results are similar for IRAS 25 and 60 $\mu$m flux densities (not
  shown). For very red sources, IRAS [12]$-$[25] $>$~1, only the MSX
  [C]$-$[E] color is closely related to the IRAS [12]$-$[25] color.
\label{f4}
}
\end{figure}

\subsection{Flux density and color definitions}\label{defs}

To be consistent with the original IRAS 2CD nomenclature we follow the
(uncorrected) \emph{color magnitude} definition used by
\citet{vdveen88}, i.e.\ color magnitude $[\alpha] - [\beta] = -2.5
\times {\log}_{10} ({\rm F}_{\alpha}/{\rm F}_{\beta}$) where ${\rm
F}_{\chi}$ is the measured flux density in Jansky (Jy) at waveband $\chi$
taken from the IRAS PSC/FSC version 2.0 or from the MSX PSC version
2.3 \citep{beichman88,egan03}. For IRAS wavebands we will use the numbers
12, 25 and 60 to designate the waveband; the center wavelength in
$\mu$m of the filter passband. For MSX we will use the capital letters A, C,
D and E for the wavebands centered at 8.28, 12.13, 14.65 and 21.3
$\mu$m respectively. If we call a source ``blue'' or ``bluer'' we
refer to the smaller or more negative value(s) of a color; relatively
larger or more positive color values are ``red'' or ``redder''. We
also use the IRAS \emph{color regions} determined by \citet{vdveen88}.
That is, statistically, the oxygen-rich stars are
located in the IRAS color regions identified with capital Roman
numerals I, II, IIIa, IIIb, IV and V along the CSE-sequence, and 
carbon-rich stars typically are located in the IRAS color regions VIa,
VIb and VII (Fig.~\ref{f2}). Color region VIII contains a complex
mixture of sources. For the remainder of this paper we will
concentrate on the oxygen-rich CSE-sequence I--V, completely ignore the 
88 sources in color region VIII, and only occasionally comment on the
carbon-rich sources in color regions VIa, VIb and VII. Later we will
also introduce MSX color regions with non-capitalized Roman numerals
(\emph{i}, \emph{ii}, \emph{iiia}, \emph{iiib}, \emph{iv} and
\emph{v}) (to keep them recognizable as distinct MSX, rather than IRAS 
color regions), but where the same type of sources can be found in the 
different 2CDs (Sect.~\ref{msx2cd}).

In our analysis we have chosen not to apply any zero-magnitude or color
corrections. One of the strengths of the IRAS 2CD as defined by
\citet{vdveen88} is that it also did not have any of these corrections
applied. The analysis of \citet{vdveen88} was therefore independent of 
different choices of extinction curves and spectral energy distributions 
(SEDs) of individual objects. Only in Sect.~\ref{coline} we will temporarily
divert and apply a zero-magnitude correction -- a simple axis shift --
to compare our data with other data sets. The zero-magnitude correction is
applied by, e.g., adding $2.5 \times {\log}_{10} (\frac{58.49}{18.29})
= 1.2622$ to the uncorrected [A]$-$[D] color and $2.5 \times
{\log}_{10} (\frac{18.29}{8.80}) = 0.7943$ to the uncorrected
[D]$-$[E] color, where 58.49, 26.51, 18.29 and 8.80 Jy are the
zero-magnitude flux densities for the MSX bands A, C, D and E,
respectively.

Finally, although we made an effort to exclude data that is ``confused''
(i.e., where the IRAS beam would include emission from several objects 
as compared to emission from a single object in the smaller MSX
beam), it is quite possible that a few sources are confused or
mis-identified. However, from the tail of Fig.~\ref{f1} we estimate
the number of mis-identifications or confused sources is less than 10\%. 
Since we are making a statistical comparison of the colors, we
expect no major difference in our conclusions when a few misidentified
objects remain in the sample.

\begin{table}
\begin{center}\caption{Least square fits$^a$ of IRAS [12]$-$[25] versus MSX
  colors}\label{t1}\begin{tabular}{lrrr|lrrr}
MSX & slope & offset & RMS  & MSX & slope & offset & RMS \\
\hline
\hline
\mbox{[A]$-$[C]}&0.468&0.230&0.224 & \mbox{[C]$-$[D]}&0.375&$-$0.099&0.154 \\
\mbox{[A]$-$[D]}&0.852&0.133&0.283 & \mbox{[C]$-$[E]}&0.765&$-$0.100&0.177 \\
\mbox{[A]$-$[E]}&1.227&0.140&0.296 & \mbox{[D]$-$[E]}&0.384& 0.003  &0.205 \\
\hline
\multicolumn{8}{l}{$^a$: MSX color = slope $\cdot$ (IRAS [12]$-$[25]) $+$ offset}\\
\end{tabular}
\end{center}
\end{table}

\section{From IRAS Colors Toward MSX Colors}\label{iras2msx}

\subsection{The IRAS CSE-sequence in MSX Colors}\label{iras2cd}

The CSE-sequence of oxygen-rich objects in the IRAS 2CD (see Fig.~\ref{f2}) 
is an exponential function of the IRAS [12]$-$[25] color resulting in an
IRAS [25]$-$[60] color. With respect to the [12]$-$[25] color region
domain the CSE-sequence overlaps in regions IIIb and IV (between $0.4 < {\rm
[12]-[25]} < 1.0$) and in regions IV and V (between $1.0 < {\rm
[12]-[25]} < 1.4$).  However, this overlap is much less than in the
[25]$-$[60] color region domain, where regions I--IIIb overlap between
$-2.4 < {\rm [25]-[60]} < -1.6$. It is therefore convenient to regard
the CSE-sequence in terms of IRAS [12]$-$[25] color. Using the
[12]$-$[25] colors at this stage does not distinguish between
carbon-rich objects in regions VIa, VIb and VII and oxygen-rich
objects along the CSE-sequence for [12]$-$[25] $<$ 0.4; we will
discuss this further in Sect.~\ref{coline}.

Strictly speaking, in terms of the center wavelength of the
wavebands, the IRAS [12]$-$[25] color would correspond roughly to the
MSX [C]$-$[E] color, with no corresponding color in the Spitzer
(GLIMPSE+MIPS) nor in the ISOGAL surveys. However, as readily
explained in \citet{ortiz05}, the MSX A filter covers the 9.7 $\mu$m
silicate feature, whereas the 11.3 $\mu$m SiC feature would appear in
the MSX C filter; the IRAS 12 $\mu$m band filter includes both
features\footnote{The Spitzer InfraRed Array Camera (IRAC) 8.0 $\mu$m
filter excludes both, which could complicate recognizing the
CSE-sequence using IRAC data.}. As the silicate feature is typically
the most prominent in the IRAS LRS, the IRAS [12]$-$[25] color may
also, or better, correspond to the MSX [A]$-$[E] color.
Figures~\ref{f3}~\&~\ref{f4} and Table~\ref{t1} compare the IRAS
colors to the MSX colors that could be derived from the
cross-identifications. This comparison shows that the IRAS [12]$-$[25] 
color can generally be converted to MSX colors, most notably in the MSX
[A]$-$[E] and [C]$-$[E] colors, but also with a reasonable color range
in the MSX [A]$-$[D] color (Table~\ref{t2}). We note that there is no
dependence on total flux density at any waveband (Fig.~\ref{f4}).
Except for the [C]$-$[E] color, the dispersion and deviation from the
trend is in general larger for very red sources, where [12]$-$[25] $>$
1, but overall the agreement is well within a third of a
magnitude. We note that most of these sources are intrinsically variable
evolved stars for which a difference in color due to non-simultaneous
observations within a mission, or observing different stellar phases
between the IRAS and MSX missions is reflected in this dispersion
\citep{robitaille07}. No conversion can be made for the IRAS
[25]$-$[60] color. This is not surprising as all MSX colors are
determined for the SED shortward of 25 $\mu$m, and are largely
insensitive to the SED and IRAS color longward of 25 $\mu$m
wavelength.

For the oxygen-rich sources along the CSE-sequence, this close
correspondence to the IRAS [12]$-$[25] color to three MSX colors is
fortunate, as the CSE-sequence thus can be represented by the IRAS
[12]$-$[25] color, and also by the MSX [A]$-$[E], [C]$-$[E] and
[A]$-$[D] colors. We will thus assume that the
  small inaccuracies of converting IRAS colors using the
linear relations in Table~\ref{t1} remain within the noise of
the whole sample of derived MSX colors. Then the IRAS CSE-sequence can be described as MSX
CSE-sequences by substituting the IRAS [12]$-$[25] color with two
different MSX colors in Eq.~2 of \citet{vdveen88} for the range of valid IRAS [12]$-$[25]
colors ($-$1.5 $<$ [12]$-$[25] $<$ $+$1.3). It follows that 
the IRAS [25]$-$[60] color can be eliminated by equating the right hand
sides of the two equations obtained above.
Rearrangement of the resulting equation yields the MSX CSE-sequences
given in Table~\ref{t3} (Sect.~\ref{msxcse}).

\begin{table}
\begin{center}\caption{Least square fits of IRAS/MSX color differences
}\label{t2}\begin{tabular}{lrrr}
IRAS $-$ MSX colors & slope & offset & RMS \\
\hline
\hline
\mbox{(IRAS [12]$-$[25]) $-$ (MSX [A]$-$[D])} & 0.148  &$-$0.133& 0.283 \\
\mbox{(IRAS [12]$-$[25]) $-$ (MSX [A]$-$[E])} &$-$0.227&$-$0.140& 0.296 \\
\mbox{(IRAS [12]$-$[25]) $-$ (MSX [C]$-$[E])} & 0.235  & 0.100  & 0.177 \\
\hline
\end{tabular}\end{center}
\end{table}

\subsection{Carbon-rich or oxygen-rich? The MSX ``OC-index''}\label{coline}

In a statistical sense IRAS colors can be used to separate the
carbon-rich sources from the oxygen-rich sources, as the carbon-rich 
CSEs typically reside in IRAS regions VIa, VIb and VII. However, when 
only MSX colors are available the separation of carbon-rich 
from oxygen-rich envelopes becomes more of a problem. Although other 
information is sometimes available upon which a clearer distinction 
between carbon-rich and oxygen-rich CSEs can be made,
(e.g.\ $J$, $K$ or $L$ band 2MASS or DENIS flux densities,
\citet{suh01,lumsden02,suh07}), the challenge is to make such
a distinction using only the data available within the MSX catalog 
\citep[e.g.,][]{lumsden02,sevenster02,ortiz05}. In our view the
most promising case is the data in Fig.~3 of \citet{ortiz05} 
which implicitly suggest that a first
separation can be made. \citet{ortiz05} show that discriminating between 
C and O rich stars is better done using $K-[A]$ versus $[A]-[D]$ colors
(their Fig.~2), but here we only use MSX colors and turn to their analysis of
 the zero-magnitude
corrected MSX $[D]-[E]$ versus MSX $[A]-[D]$ color (their Fig~3). The
samples in \citet{ortiz05} are well defined as the data from
\citet{loup93} that they use are identified as carbon-rich stars and
indeed typically, but not uniquely, populate IRAS color regions VIa,
VIb and VII \citep[][their Fig.~3]{loup93}. Similarly, but vice versa
for oxygen-rich stars in regions I--V, this reminds one that the IRAS
boundary between carbon-rich and oxygen-rich color regions is not
strict.

\begin{figure} \begin{center}
\plotone{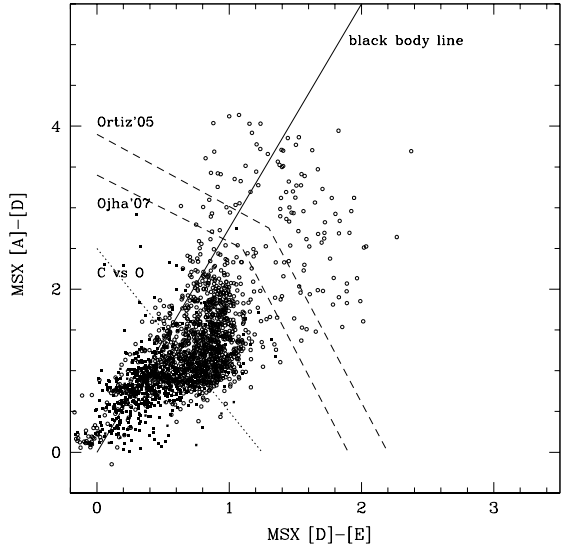}
\end{center}
\caption{\small Zero-magnitude corrected MSX $[D]-[E]$ versus
  $[A]-[D]$ 2CD following \citet{ortiz05} \citep[and][]{ojha07} for
  the final selection of 2455 sources after removing those in IRAS
  color region VIII (88 sources). For 544 sources data could not be
  plotted in this diagram (see text). Filled squares are sources from IRAS
  color regions VIa, VIb and VII, the supposedly carbon-rich sources;
  open circles are supposedly oxygen-rich sources from IRAS color
  regions I--V. The dotted line labeled ``C vs O'' ($[A]-[D] = - 2
  \cdot ([D]-[E]) + 2.5$) is an empirical line derived from Fig.~3 in
  \citet{ortiz05}. It suggests a division between carbon-rich and oxygen-rich
  sources in their data (see text and Fig.~\ref{f6}).  Also drawn
  (dashed) are the lines from \citet{ortiz05} and \citet{ojha07} that
  depict a gap between stars and planetary nebulae or transition
  objects in their data; the gap is also apparent here.  The (solid)
  black body temperature line ranges from 10,000 K at the origin to
  about 140 K at $[D]-[E] = 2$ (note that the black body temperature
  scale in \citet{ortiz05} is in error [R.~Ortiz, priv.\ comm.]),
  crossing the ``C vs O'', ``Ojha07''and ``Ortiz05'' line at T $\sim$
  400, 255 and 230 K respectively. These lines are roughly
  perpendicular to the black body temperature line and thus may be
  interpreted as temperature separators.
\label{f5}
}
\end{figure}

In Fig.~3 of \citet{ortiz05} the carbon-rich stars (filled squares)
are mostly separated from the oxygen-rich OH/IR objects by the
line $[A]-[D] = - 2 \cdot ([D]-[E]) + 2.5$ using zero-magnitude
corrected colors. This separator is the dotted line in our Fig.~\ref{f5} 
and labeled ``C vs O''\footnote{As an interesting side note, these
are colors involving the MSX D filter, whereas in the previous section
(\ref{iras2cd}) we mentioned that the ``carbon feature'', the SiC
line, appears in the MSX C filter. Plausible components with carbon-bearing
molecules which could affect the flux density in the D filter are those of HCN
and C$_2$H$_2$ at $\sim$ 14 $\mu$m \citep[see e.g.,][]{blommaert05}. }. Is it possible to use this empirical line to
generally make the division between carbon-rich and oxygen-rich
objects when the composition of the envelope is not known a-priori, 
e.g., such as in samples used in \citet{ojha07} and here?

\begin{figure} \begin{center}
\plotone{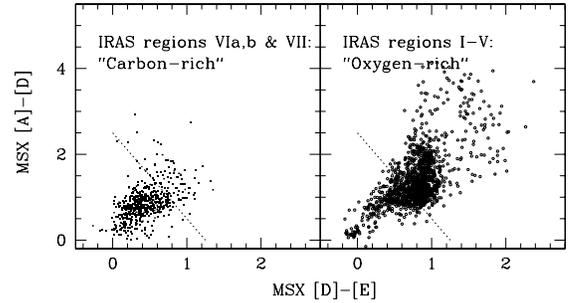}
\end{center}
\caption{Zero-magnitude corrected MSX $[D]-[E]$ versus
  $[A]-[D]$ 2CD as Fig.~\ref{f5}, but now separated for IRAS color
  regions VIa, VIb and VII (``Carbon-rich''; {585 sources for which an
  ``OC-index'' could be calculated out of 966 sources in the
  carbon-rich IRAS regions VIa, VIb and VII) versus IRAS
  color regions I--V (``Oxygen-rich''; 1326 of 1489 sources). The (dotted)
  line is the empirical ``C vs O'' line from Fig.~\ref{f5}. Selecting
  sources to the lower left of the ``C vs O'' line, i.e., selecting
  sources for which the ``OC-index'' (Eq.~\ref{eq:e1}) is less than
  zero (see text; Sect.~\ref{coline}), helps to increase the fraction
  of carbon-rich sources taken from the sample. However, selecting
  sources to the upper right, with a positive ``OC-index'', very
  effectively deselects oxygen-rich sources with T $>$ 400 K (see
  Fig.~\ref{f7}).}
\label{f6}
}
\end{figure}

In Fig.~\ref{f5} the sources of our sample in IRAS color regions VIa,
VIb and VII (filled squares) and the other regions (open circles)
excluding region VIII (88 sources) seem to overlap in the densely
populated part of the figure around the ``C vs O'' line. 
Note that only 1911 of the original 2455 sources could be plotted in
this figure, as 544 sources are missing at least one good quality A, D, 
or E flux measurement. Figure~\ref{f6}
separates the two samples to show that supposed carbon-rich 
sources from IRAS region VIa, VIb and VII do appear in the 
oxygen-rich (top right) side of the ``C vs O'' dotted line. Supposed
oxygen-rich sources also appear on the carbon-rich (bottom left)
side. The IRAS color regions VIb and VII as described by
\citet{vdveen88} may still contain a large fraction of oxygen-rich
stars based on the percentage of LRS classes 21--29 and 14--16
\citep{olnon86} reported in the samples \citep{vdveen88}.  There is
also some fraction of supposed oxygen-rich stars in regions II--IV which
have LRS classes consistent with carbon-rich stars.  This may also be seen, e.g.,
in \citet{suh01} where evolutionary models of carbon-rich stars do
traverse oxygen-rich IRAS regions IIIa and IIIb, and models for
oxygen-rich stars border region VII.

Not surprisingly this means that the IRAS region division between oxygen-rich
and carbon-rich stars is not
perfect. We therefore investigated whether a better separation between
carbon-rich and oxygen-rich objects can be made using the MSX [D]$-$[E] versus
[A]$-$[D] colors dividing at the ``C vs O'' line, with the
understanding that the MSX separation may also not be perfect.  We
define the MSX ``OC-index'' for sources with good quality MSX A, D and
E flux densities:
\begin{equation}\label{eq:e1}
 {\rm OC-index}\ =\ ([A]-[D])\ +\ 2\cdot([D]-[E])\ +\ \zeta\ ,
\end{equation} with
$\zeta = -$2.5 for zero-magnitude corrected MSX colors and $\zeta =
+$0.35 for uncorrected MSX colors. The OC-index is thus negative
if the source is on the supposedly carbon-rich side of the ``C vs O''
line and positive on the supposedly oxygen-rich side.

\begin{figure} \begin{center}
\plotone{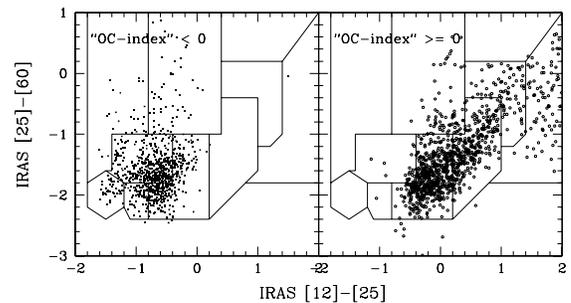}
\end{center}
\caption{IRAS 2CDs of the 1911 sources separated according to the
  OC-index (``C vs O'' line) for potentially carbon-rich objects
  (\emph{left}; 886 sources) and oxygen-rich objects (\emph{right};
  1025 sources). In general the division using the OC-index puts the
  sources in the expected IRAS regions, but the OC-index is not a good divider
  for sources hotter than 400 K (where the ``C vs O'' line crosses the
  black body temperature line in Fig.~\ref{f5}) as no oxygen-rich
  objects end up in IRAS regions I and II.  }
\label{f7}
\end{figure}

Figure~\ref{f7} shows the IRAS 2CD for the sample,
divided according to the OC-index separately. Although the OC-index 
seems to generally divide the carbon-rich from the oxygen-rich objects,
it mistakenly identifies potential oxygen-rich sources in IRAS regions I and
II as carbon rich. It does select the cooler and thicker CSEs in IRAS color
regions IIIa--V, but since we are interested in the full CSE-sequence,
(including IRAS color regions I and II), using this division does not 
suit our needs in this context. We conclude that regardless of the potential 
to increase the fraction of carbon-rich sources in samples of hot objects (T $>$ 400
K, left of the border between region VII and IIIa), the OC-index is
more a measure of temperature than CSE chemistry (Fig.~\ref{f7}).

\begin{figure*} \begin{center}
\plotone{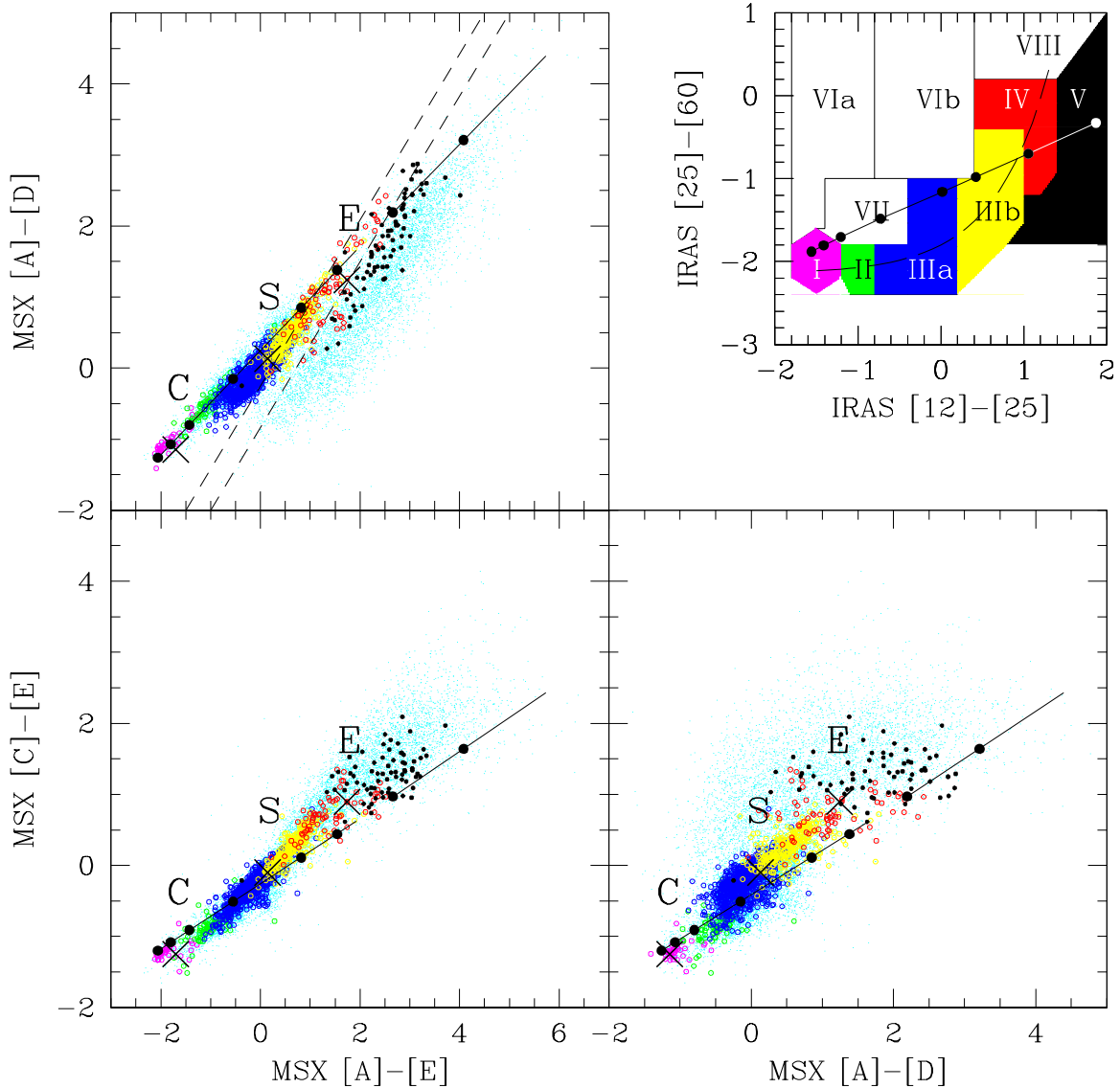}
\end{center}
\caption{\small \emph{Top right:} IRAS 2CD with color coded regions along the
  CSE-sequence. These color codes are used for the color of each
  source in the other panels, in the MSX 2CDs. For
  example, sources with IRAS colors consistent with being located
  in IRAS color region IIIb (yellow) appear in yellow in the MSX 2CDs.
The black body
  temperature curve temperature labels are omitted at the ticks for
  clarity; they are the same as in Fig.~\ref{f2}, i.e., 10,000 K,
  2000, 1000, 500, 300, 250, 200 and 160 K from left to right
  respectively. \emph{Other panels:} MSX 2CDs of the 1489 sources in
  IRAS regions I--V. The color of the source depicts in which IRAS
  region it resides (see top right panel). Small cyan (lighter blue)
  points are sources with good MSX colors from the entire MSX catalog.
  The IRAS matches discussed here are a subset of mainly brighter
  sources which are in the Galactic plane. The CSE-sequence is the
  line connecting the crosses labeled C, S
  and E (the line itself is not drawn, for clarity), plotted at [12]$-$[25] = $-$1.5, 0 and $+$1.3
  respectively. The solid line is the black body temperature curve
  starting at 10,000 K in the lower left, interrupted between 230 K
  and 200 K for reference, and ending at 130 K.
The fat-dot ticks are at the same temperatures as in the IRAS 2CD. 
The dashed lines in
  the top left panel depict the lines separating sources from IRAS
  region IIIb from those from IRAS region IV and sources from region
  IV from those from IRAS region V -- see text. }
\label{f8}
\end{figure*}

The main interest in the present study is in the oxygen-rich sources 
along the CSE-sequence. As none of the MSX colors relates closely to 
the IRAS [25]$-$[60] color,
it is expected that the carbon-rich and oxygen-rich sources overlap in
MSX colors where the uncorrected IRAS [12]$-$[25] color is less than
$\sim$ +0.4, the farthest extent of IRAS color region VIb. To
eliminate the pollution of an oxygen-rich sample along the
CSE-sequence with likely carbon-rich sources in the MSX color samples,
we therefore removed from our sample all sources from the carbon-rich IRAS color
regions VIa, VIb and VII (and VIII). However, we did not remove the
388 sources with OC-index $<$ 0 from the sample as this would
eliminate \emph{all} hot (T $>$ 400 K) oxygen-rich objects as well.
Thus a reduced sample of 1489 sources with good MSX colors in the 
oxygen-rich IRAS regions was used in the remainder of this paper.
Note that we chose to eliminate the likely
carbon-rich sources. An option to introduce different symbol colors
(see Fig.~\ref{f8}) would not add to clarity in the figures later on,
whereas opting for symbol colors according to their IRAS [12]$-$[25]
color, and thus mislabeling them as oxygen-rich sources, would just
plot them on top of the distribution of oxygen-rich sources with the
same symbol color. That is, leaving them out from the comparison of
infrared colors of oxygen-rich sources seems proper and does not alter
our conclusions.

\begin{figure} \begin{center}
\plotone{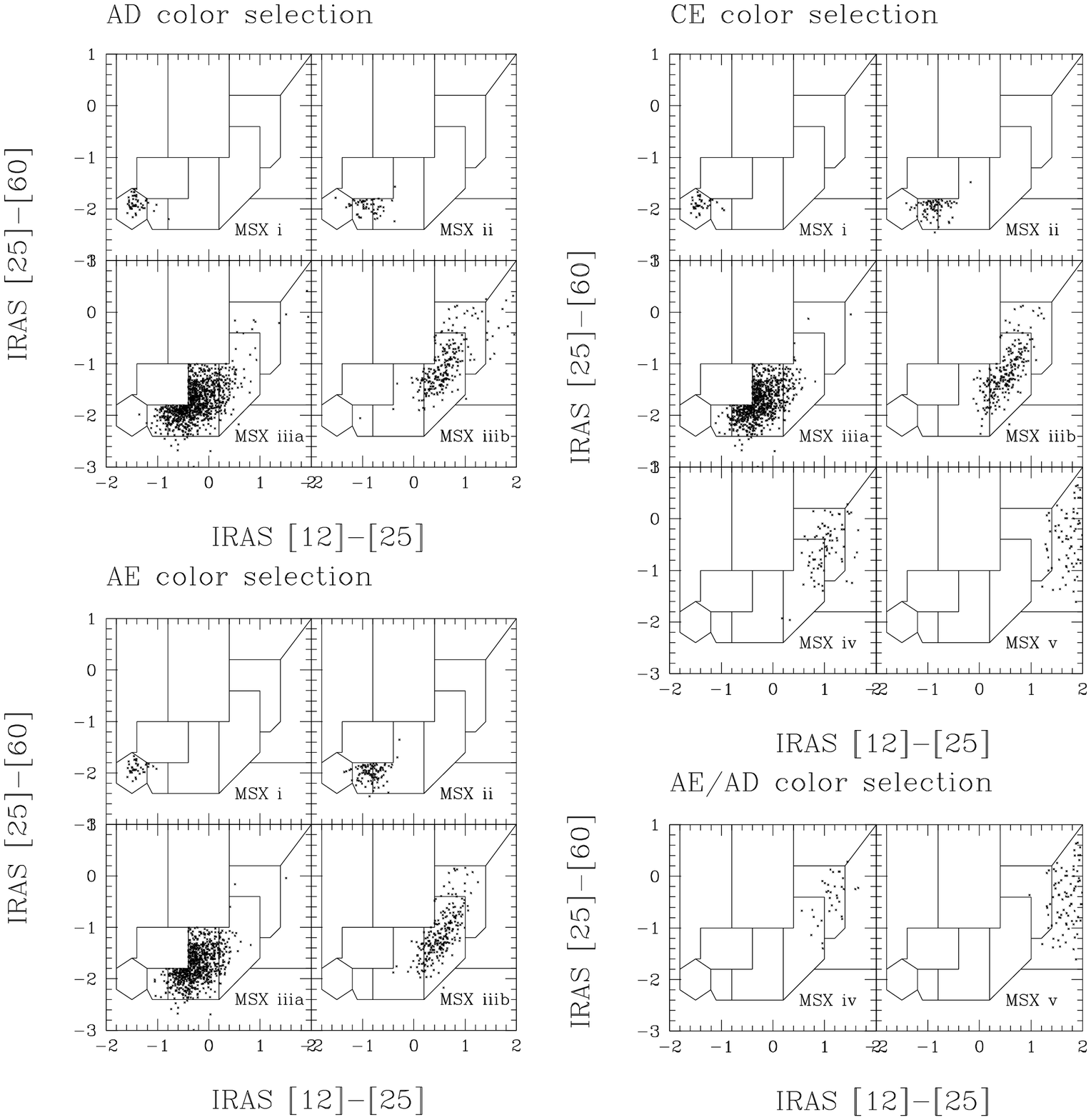}
\end{center}
\caption{Resulting IRAS 2CD when the selection for MSX color regions
  is applied. IRAS color regions I, II and IIIa can readily be
  selected using a single MSX color, [A]$-$[D], [A]$-$[E] or
  [C]$-$[E]. Because the MSX [C]$-$[E] color corresponds best to the
  IRAS [12]$-$[25] color, IRAS color regions IIIb, IV and V can also
  be selected using the [C]$-$[E] color. If the [C]$-$[E] color is
  unavailable, due to a missing or unreliable C filter flux density,
  these regions can only be selected using a combination of the
  [A]$-$[D] and [A]$-$[E] colors. }
\label{f9}
\end{figure}

\section{MSX Two-color Diagrams}\label{msx2cd}

Using the sample of 1489 generally oxygen-rich sources, MSX 2CDs were
constructed. In an attempt to describe MSX regions that would
correspond to IRAS regions, it is important to determine MSX 2CDs in
which the different IRAS regions clearly separate one from another.
As with the IRAS regions, some overlap of sources is expected
and unavoidable with these statistical approaches. All possible combinations
of MSX 2CDs were investigated. However, as can be deduced from the
IRAS-MSX relations in Sect.~\ref{iras2cd} and Fig.~\ref{f3}, the MSX
colors [A]$-$[D], [A]$-$[E], and [C]$-$[E], which have a slope larger than
0.5, are most useful. Furthermore, a color using the E-filter is the
most distinctive as it gives the largest color range (but note
Sect.~\ref{error}). Also, because all MSX colors represent the
IRAS [12]$-$[25] color, a more or less linear relation between any of
these two MSX colors may be expected albeit with different absolute
color ranges. In Fig.~\ref{f8} these MSX 2CDs are shown with different
source colors, depending in which IRAS color region the MSX source
resides (see top right panel). In general the trend is along the
CSE-sequence which is no surprise as the CSE-sequence, the black body
temperature curve and the IRAS [12]$-$[25] color all align with the
sequence of IRAS color regions I--V.

\subsection{MSX color regions and the CSE-sequence}\label{msxcse}

Figure~\ref{f8} shows that sources from a particular IRAS region (a
specific color) generally populate a confined region in the MSX
2CD diagrams. We therefore determined MSX 2CD color regions for
sources with IRAS colors of color region I, II, etc. We will name the
MSX region with sources from IRAS color region I, MSX color region
\emph{i} and similarly, e.g., sources from IRAS color region IIIb, MSX
color region \emph{iiib}, etc. Using our limited sample of 1489
sources total, 
we empirically determined approximate MSX color region
boundaries as follows (Fig.~\ref{f8}):

\medskip

\[ \mbox{MSX\ \ {\Large\bfseries  \emph{i$\phantom{iib}$}}\ :\ } \left\{
\begin{array}{lclcll}
-2.0\phantom{0} & \le & \mbox{[A]$-$[D]} & < & -0.9\phantom{0} & {\rm or} \\ \\
-3.0\phantom{0} & \le & \mbox{[A]$-$[E]} & < & -1.5\phantom{0} & {\rm or} \\ \\
-2.0\phantom{0} & \le & \mbox{[C]$-$[E]} & < & -1.1\phantom{0} 
\end{array}
\right . \]

\medskip

\[ \mbox{MSX\ \ {\Large\bfseries  \emph{ii$\phantom{ib}$}}\ :\ } \left\{
\begin{array}{lclcll}
-0.9\phantom{0} & \le & \mbox{[A]$-$[D]} & < & -0.6\phantom{0} & {\rm or} \\ \\
-1.5\phantom{0} & \le & \mbox{[A]$-$[E]} & < & -0.7\phantom{0} & {\rm or} \\ \\
-1.1\phantom{0} & \le & \mbox{[C]$-$[E]} & < & -0.75 
\end{array} \right . \]

\medskip

\[ \mbox{MSX\ \ {\Large\bfseries  \emph{iiia}}\ :\ } \left\{
\begin{array}{lclcll}
-0.6\phantom{0} & \le & \mbox{[A]$-$[D]} & < & +0.4\phantom{0} & {\rm or} \\ \\
-0.7\phantom{0} & \le & \mbox{[A]$-$[E]} & < & +0.4\phantom{0} & {\rm or} \\ \\
-0.75 & \le & \mbox{[C]$-$[E]} & < & +0.1\phantom{0} 
\end{array} \right . \]

\medskip

\[ \mbox{MSX\ \ {\Large\bfseries  \emph{iiib}}\ :\ } \left\{
\begin{array}{lclcll}
+0.4\phantom{0} & \le & \mbox{[A]$-$[D]} & < & +5.0\phantom{0} & {\rm and} \\
+0.1\phantom{0} & \le & \mbox{[C]$-$[E]} & < & +0.55 & {\rm or} \\ \\
\multicolumn{5}{l}{\mbox{1.4$\cdot$([A]$-$[E]) $-$ ([A]$-$[D])} <  +0.8} & {\rm and} \\
+0.4\phantom{0} & \le & \mbox{[A]$-$[E]} & < & +1.1\phantom{0} & {\rm or} \\ \\
+0.1\phantom{0} & \le & \mbox{[C]$-$[E]} & < & +0.55 
\end{array} \right . \]

\medskip

\[ \mbox{MSX\ \ {\Large\bfseries  \emph{iv$\phantom{i}$}}\ :\ } \left\{
\begin{array}{lclcll}
\multicolumn{5}{r}{+0.8 \le \mbox{1.4$\cdot$([A]$-$[E]) $-$ ([A]$-$[D])} < +1.3} & {\rm and} \\
+0.4\phantom{0} & \le & \mbox{[A]$-$[E]} & < & +7.0 & {\rm or} \\ \\
+0.55 & \le & \mbox{[C]$-$[E]} & < & +1.0\phantom{0} 
\end{array} \right . \]

\medskip

\[ \mbox{MSX\ \ {\Large\bfseries  \emph{v$\phantom{iii}$}}\ :\ } \left\{
\begin{array}{lclcll}
\multicolumn{5}{l}{\mbox{1.4$\cdot$([A]$-$[E]) $-$ ([A]$-$[D])} \ge +1.3 } & {\rm and} \\
+1.1\phantom{0} & \le & \mbox{[A]$-$[E]} & < & +7.0 & {\rm or} \\ \\
+1.0\phantom{0} & \le & \mbox{[C]$-$[E]} & < & +5.0\phantom{0} &  
\end{array} \right . \]

\medskip

Two regions in the [A]$-$[E] versus [A]$-$[D] diagram
(upper left panel in Fig.~\ref{f8}) are not
accounted for in the above determination due to the lack of sources. 
The first region is for 1.1 $<$ [A]$-$[E] $<$ 7.0 and with 1.4$\cdot$([A]$-$[E])
$-$ ([A]$-$[D]) $<$ +0.8, where sources from either IRAS region IIIb and
IV can be expected. Sources from either IRAS region IV and V can be
expected between +0.4 $\le$ [A]$-$[E] $<$ 1.1 and with
1.4$\cdot$([A]$-$[E]) $-$ ([A]$-$[D]) $>$ +1.3.

We also included the beginning, zero and end points of the CSE-sequence in
the MSX 2CDs. These are the locations of the crosses in the MSX 2CDs
(Fig.~\ref{f8}) where IRAS [12]$-$[25] equals to $-$1.5 (labeled C),
zero (labeled S) and $+$1.3 (labeled E). The full CSE-sequence is a
straight line between these points and can be obtained by eliminating
the [25]$-$[60] color from the CSE-sequence equation in \citet[][their
Eq.~2]{vdveen88}, or simply by equating the expressions for
[12]$-$[25] from Table~\ref{t1} for different MSX colors over the
[12]$-$[25] range in MSX colors where the CSE-sequence is defined. The
expressions and color ranges for the CSE-sequence in MSX colors
adapted here are given in Table~\ref{t3}. It can be
seen that indeed the CSE-sequence starts in MSX color region \emph{i},
has its zero point close to the border between regions \emph{iiia} and
\emph{iiib}\footnote{In Fig.~\ref{f8} the yellow symbols of region
\emph{iiib} overlap and unfortunately \emph{overplot} the blue symbols
of region \emph{iiia}, making it unclear that the blue
symbols continue under the yellow symbols. However, note that the yellow
symbols do not extend much past the negative (lower left) side of
[A]$-$[E] = 0.}, and ends in region \emph{iv}. It does not enter the
region (\emph{v}) with the reddest sources, which are mostly (proto-)planetary
nebulae.

To obtain a subsample of sources that is characterized by colors of a
certain IRAS color region, the selection criteria for the same (lower
case) MSX color region can be directly applied if the parent sample
contains uncorrected MSX colors. For example, a
  subsample of sources with CSEs resembling the IRAS IIIa region
  colors (blue in Fig.~\ref{f8}) can be drawn directly from the MSX
  catalog by applying one or more of the three equations for MSX
  region \emph{iiia}.
On the other hand, to obtain such a subsample from a parent
sample that contains zero-magnitude corrected MSX colors,
one would have to
subtract the zero-flux magnitude fractions from Sect.~\ref{defs} from
the MSX source colors first\footnote{Because
the MSX A and D waveband filters are similar to the ISOGAL 7 and 15
$\mu$m waveband filters, selection on the zero-magnitude corrected
ISOGAL [7]$-$[15] color ranges may appear to be similar to selection
on the zero-magnitude color corrected MSX [A]$-$[D] color
\citep{messineo04}.}. As a reminder, we emphasize that carbon-rich
sources in IRAS regions VIa, VIb and VII will overlap with MSX color
regions \emph{i}, \emph{ii} and \emph{iiia}, with a small tail in
\emph{iiib}--\emph{v}. Using only MSX colors it is impossible to make
an accurate estimate of the fraction of 
contamination with carbon-rich sources in a
sample. However, to get an indication, in 
Table~\ref{t4} we show the number of
sources from the cross-identified IRAS/MSX data base (2543 sources;
Sect.~\ref{select}) 
in each IRAS color region, when sources
are selected using any of the MSX selection rules
above. It is apparent from this table that
  selecting a sample using the MSX color relations in general selects
  the sources with similar IRAS colors (e.g., MSX \emph{iiia} and IRAS
IIIa), but that also sources from nearby IRAS regions may fulfill the
MSX selection rule. This, in a statistical approach, is not worrisome;
e.g., using the MSX [A]$-$[D] color to select MSX \emph{iiia} sources
will yield 82\% (816/995) of IRAS IIIa sources, provided that the
sample is known to be mostly oxygen-rich along the CSE-sequence. 
Because of the degeneracy in the vertical axis of the IRAS 2CD when
only MSX colors (or IRAS [12]$-$[25] colors) are available, the
carbon-rich objects in IRAS regions VIa, VIb and VII also fulfill the
MSX selection rules. The approximate number of sources in these IRAS
regions (according to increasing [12]$-$[25] color) is also given in
Table~\ref{t4} per MSX selection rule.
If the
degeneracy between oxygen-rich and carbon-rich cannot be resolved,
e.g. with the IRAS [25]$-$[60] color, then the \emph{pure} fraction of IRAS
region IIIa sources of the MSX \emph{iiia} sources using the [A]$-$[D]
color selection degrades to 50\% (816/1624). Nevertheless, selecting
on e.g. MSX \emph{iiia} will thus indeed mostly yield IRAS IIIa
sources, partly because some of the IRAS VIa, VIb and VII sources may be
actually oxygen-rich since the IRAS boundaries also do not exactly
separate the two populations.
 Because the density of carbon-rich sources is highest for
[12]$-$[25] $<$ 0, the highest contamination (cq.\ overlap) is in MSX
color regions \emph{i}, \emph{ii} and \emph{iiia}. For these regions
other selection criteria to increase the oxygen-rich fraction
are advisable (e.g.\ \citet{messineo02},
Sect.~\ref{sio86}). Again, the oxygen-rich and carbon-rich objects
overlap around the dividing regions in the IRAS 2CD, in particular
near IRAS region VII, so the numbers in
Table~\ref{t4} are indicative
values only for any sample other than this IRAS/MSX
  cross correlated data set. For the particular example above, selecting MSX 
\emph{iiia} sources using the [A]$-$[D] color, the fraction of oxygen-rich
objects is 61\% (995/1624). That is, for such a sample the remainder, 39\%,
may be carbon-rich and thus will not harbor SiO masers to begin with 
(Sect.~\ref{sio86}).

\begin{table}
\begin{center}\caption{CSE-sequence in MSX two-color diagrams
  (Fig.~\ref{f8})}\label{t3}
\begin{tabular}{c|c}
MSX CSE-sequence relations & Valid MSX CSE color ranges \\
\hline
\hline
\mbox{[A]$-$[D]} = 0.694 $\cdot$ (\mbox{[A]$-$[E]}) $+$ 0.036 &
$-$1.145 \mbox{$<$ [A]$-$[D] $<$} $+$1.241 \\
\mbox{[C]$-$[E]} = 0.623 $\cdot$ (\mbox{[A]$-$[E]}) $-$ 0.187 &
$-$1.701 \mbox{$<$ [A]$-$[E] $<$} $+$1.735 \\
\mbox{[C]$-$[E]} = 0.898 $\cdot$ (\mbox{[A]$-$[D]}) $-$ 0.219 &
$-$1.248 \mbox{$<$ [C]$-$[E] $<$} $+$0.895 \\
\hline
\end{tabular}\end{center}
\end{table}

\subsubsection{Possible error in MSX PSC 2.3 ?}\label{error}

In our investigation we noticed possible error in the MSX PSC
version 2.3 \citep{egan03}.  In the MSX [A]$-$[E] versus [A]$-$[D] 2CD 
(the top left panel in Fig.~\ref{f8}) there seems to be a different 
distribution of sources which is shifted to the right and away from the 
CSE-sequence with respect to the main-populated area when \emph{all} 
MSX sources with good quality photometry are used\footnote{This may
also be seen in the [A]$-$[D] versus [C]$-$[E] 2CD in the bottom
right panel, where many sources show up above the main-populated
area.}. We could not 
account for this different distribution except by positing a somewhat 
different, redder population.  In order to help confirm this possibility, 
we made the same 2CD diagram with data retrieved from MSX PSC version 
1.2 \citep{egan99}. When compared to the PSC version 2.3 2CD there seems 
to be a more continuous distribution of redder (and weaker) objects in 
the diagram made from version 1.2 data (not shown).

From both versions of the PSC we selected objects with quality 3 or 4 in A, D,
and E and constructed each color from the catalog data. Typically the
sources weaker in flux density in the MSX A filter are affected, with
a shift in [A]$-$[E] of $\sim$ 0.5--1 magnitude redward.  We
understand that the difference between PSC versions 1.2 and 2.3 is
that the photometry has been improved, and weaker sources have been
included. However, we would not have expected the ``gap'' in the
distribution of colors in the 2.3 data (top left panel in
Fig.~\ref{f8}) that does not appear in the 1.2 data. The reason for
this is currently unknown, but seems to originate from MSX E filter
(21.3 $\mu$m band) fluxes that are a factor of about two too high for
most of the weaker A filter (8.28 $\mu$m band)
sources\footnote{A referee has explained the
  probable cause to be a combination of uncompensated variable
  response during the mission and the different techniques used for
  the extraction of the photometry in the different versions.}.


\begin{table*}
\caption{Correlation matrix for sources for which the
    MSX color could be calculated and fall within the ranges of the
    equations given in Sect.~\ref{msxcse} ("Total"). The numbers are the source
    counts in each IRAS color region for each of the MSX color
    selection rules, and indicate the potential contamination level of
    other sources than targeted using the MSX rules.}\label{t4}
\hspace*{-1.5cm}
\begin{tabular}{l|rrrrrr|rrrr|rr}
\\
MSX color&\multicolumn{6}{c}{IRAS CSE color
  region}&\multicolumn{4}{c}{Non-CSE IRAS region}&\multicolumn{2}{c}{Total}\\
selection used & I & II & IIIa & IIIb & IV & V & VIa & VII & VIb & VIII & I--V (O-rich) & I--VIII\\
\hline
\hline\smallskip
MSX \emph{i} using: \\
\mbox{[A]$-$[D]} & 48 (92\%/27\%) & 3 & 1 & - & - & - & 52 & 71 & 1 & - & 52 (30\%) & 176\\
\mbox{[A]$-$[E]} & 35 (90\%/41\%) & 4 & - & - & - & - & 12 & 34 & - & - & 39 (46\%) & 85\\
\mbox{[C]$-$[E]} & 35 (90\%/37\%) & 4 & - & - & - & - & 10 & 46 & - & - & 39 (41\%) & 95\\
\hline\smallskip
MSX \emph{ii} using: \\
\mbox{[A]$-$[D]} & 6 & 29 (54\%/12\%) & 19 & - & - & - & 32 & 137 & 17 & - & 54 (23\%) & 240\\
\mbox{[A]$-$[E]} & 6 & 52 (50\%/14\%) & 47 & - & - & - & 19 & 244 & 16 & - &105 (27\%) & 384\\
\mbox{[C]$-$[E]} & 6 & 48 (61\%/17\%) & 25 & - & - & - & 17 & 183 & 10 & - & 79 (27\%) & 289\\
\hline\smallskip
MSX \emph{iiia} using: \\
\mbox{[A]$-$[D]} & 2 & 58 & 816 (82\%/50\%) &110 & 6 & 3 &23 & 410 &179 &17 & 995 (61\%) & 1624\\
\mbox{[A]$-$[E]} & - & 15 & 699 (90\%/69\%) & 65 & 1 & 1 & - & 157 & 82 & - & 781 (77\%) & 1020\\
\mbox{[C]$-$[E]} & - & 19 & 713 (88\%/64\%) & 72 & 1 & 1 & 4 & 209 & 93 & 1 & 806 (72\%) & 1113\\
\hline\smallskip
MSX \emph{iiib} using: \\
\mbox{ [A]$-$[D] \& [C]$-$[E]} & - & - & 11 & 141 (83\%/75\%) & 17 & - & - & - & 5 & 15 & 169 (89\%) & 189\\
\mbox{ [A]$-$[D] \& [A]$-$[E]} & - & - & 31 & 203 (80\%/68\%) & 20 & - & - & 2 &19 & 24 & 254 (85\%) & 299\\
\mbox{[C]$-$[E]}                      & - & - & 39 & 191 (77\%/68\%) & 18 & - & - & - &14 & 19 & 248 (88\%) & 281\\
\hline\smallskip
MSX \emph{iv} using: \\
\mbox{ [A]$-$[D] \& [A]$-$[E]} & - & - & 1 & 11 & 27 (64\%/41\%) & 3 & - & - & 1 & 23 & 42 (64\%) & 66\\
\mbox{[C]$-$[E]}                      & - & - & 1 & 29 & 41 (51\%/36\%) & 9 & - & - & 1 & 32 & 80 (71\%) &113\\
\hline\smallskip
MSX \emph{v} using: \\
\mbox{ [A]$-$[D] \& [A]$-$[E]} & - & - & - & - & 9 & 72 (89\%/75\%) & - & - & - & 15 & 81 (84\%) & 96\\
\mbox{[C]$-$[E]}                      & - & - & - & - &10 & 60 (86\%/73\%) & - & 1 & - & 11 & 70 (85\%) & 82\\
\hline
\end{tabular}
\end{table*}

We have investigated whether our analysis also suffers from possible
inaccuracies or effects in the weaker 8.28 $\mu$m objects. Fortunately
the cross-identified objects, that are also detected by the less
sensitive IRAS 12 $\mu$m detector, belong to the brighter MSX sample
for which almost no sources are expected to be affected. This can also
been seen from the lack of color-coded source symbols shifted to the
right of the main distribution in the [A]$-$[E] versus [A]$-$[D]
diagram. Note that the black dots of MSX color region \emph{v} (or
IRAS color region V, i.e., mostly planetary nebula) are shifted
redward, because they are intrinsically redder than the CSE-sequence
sources.

\section{86 GHz SiO Masers and MSX Color}\label{sio86}

Previously the IRAS 2CD has been very successful in selecting objects
to target: for example, for 1612 MHz maser emission
\citep[e.g.,][]{telintel91}.  Similarly, IRAS colors have been used
to select targets for 86 GHz ($J=2\rightarrow1, v=1$) and 43
GHz ($J=1\rightarrow0, v=1 \mathrm{\ and\ } v=2$) SiO masers
\citep[e.g.,][]{haikala90,haikala94,nakada93}. \citet{jiang95} have
made a careful analysis of the 43 GHz SiO maser detection rate as function of IRAS color and IRAS color
region from the data collected by
\citet{izimiura94,izimiura95a,izimiura95b} in the Galactic bulge and
disk. The Izumiura data were selected based on color \emph{and} variability index
with likely carbon-rich stars from IRAS regions VIa, VIb and VII
removed.

In addition to an overall constant detection rate of SiO masers of
about 61$\pm$3\% as function of [12]$-$[25] color, \citet{jiang95} 
suggest that the detection rate in the Galactic bulge is higher than in 
the disk. Since in both the Galactic bulge and disk the IRAS survey suffered from 
confusion, the \citet{jiang95} study may have suffered from selection effects. 
It will be interesting to repeat this analysis based on MSX colors alone.

Another approach was used by \citet{messineo02} who bypassed the IRAS
color selection and used the combined ISOGAL-DENIS data to empirically
determine a complicated, extinction-corrected, color-magnitude-based 
selection from MSX [A]$-$[D] color and also on the MSX variability index 
in the Galactic bulge region.  A cross-identification of \citet{messineo02}'s  
target sources with the IRAS PSC (10\arcsec\ radius as in Sect.~\ref{select}) 
results in only three sources with good photometry in all three IRAS bands, and
only 81 with good photometry in [12]$-$[25] color. A selection based on IRAS 
color in this case would not have yielded an interesting sample, whereas the 
simple MSX tool presented here would have been extremely helpful in ease of use
compared to their selection.

\begin{figure} \begin{center}
\plotone{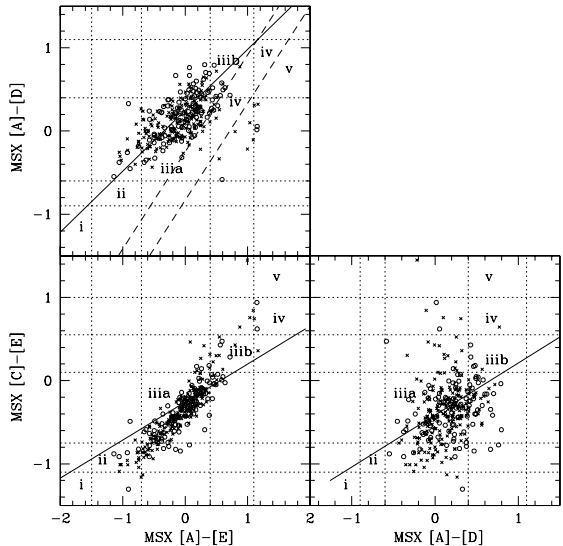}
\end{center}
\caption{MSX 2CDs for 86 GHz SiO maser targets from
  \citet{messineo02}. Crosses are detections, open circles are 
  non-detections. Most sources selected by \citet{messineo02} are
  assigned to MSX color region \emph{iiia}, comparable to the IRAS
  color region (IIIa) where for the Galactic bulge many SiO masers are
  expected \citep{jiang95}. Dotted horizontal and vertical lines show
  the MSX color region boundaries from Sect.~\ref{msxcse}. The other
  lines are as in Fig.~\ref{f8}.}
\label{f10}
\end{figure}

\begin{figure} \begin{center}
\plotone{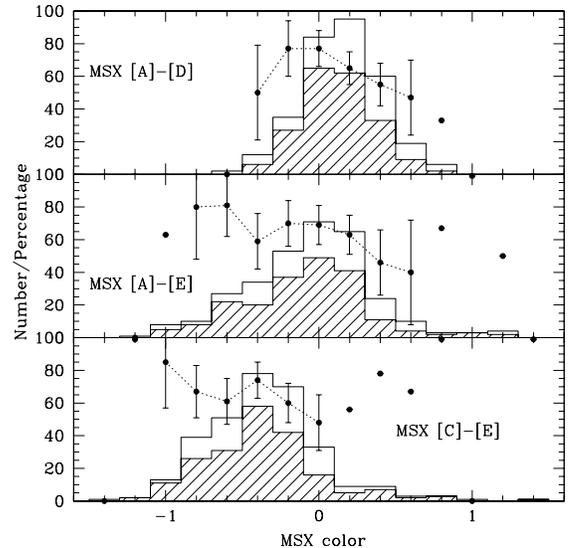}
\end{center}
\caption{Detection rates as a function of MSX color for the SiO masers
  in \citet{messineo02}. Shown are the histogram of targeted sources
  and the shaded histogram of the detected sources. The detection rate
  per 0.2 magnitude bin is depicted by dots, which are connected and
  have error bars only if the number of targets in the bin is ten or
  more. We note that the MSX photometry has changed slightly between
  PSC (preliminary) version 1.2 and version 2.3, so \citet{messineo02}'s
  strict [A]$-$[D] color selection, and the inclusion of ISOGAL selected
  sources, does not hold any more in our [A]$-$[D] color diagrams for
  some individual sources.  }
\label{f11}
\end{figure}

To assess their selection in terms of the [12]$-$[25] color analysis
by \citet{jiang95},  we plot the data from \citet{messineo02} in the MSX
2CDs in Fig.~\ref{f10} and (non)-detection histograms
in Fig.~\ref{f11}. Most sources selected by \citet{messineo02} are
located in MSX region \emph{iiia} which was the IRAS region they were
targeting. Their sample likely includes more carbon-rich sources than the
sample of \citet{jiang95} as \citet{messineo02} could not remove those sources using the 
IRAS [25]$-$[60] color.

The \citet{messineo02} 86 GHz SiO maser detection rate is consistent with an overall
constant 65-66\% as a function of either [A]$-$[D], [A]$-$[E] or
[C]$-$[E] color, i.e., with (unavailable) [12]$-$[25] color (Fig.~\ref{f11}).
Remarkably, this is the same result as \citet{jiang95} for the 43 GHz masers. 
For MSX region \emph{iiia}, where the majority of the \citet{messineo02} sample lies, 
the IRAS/MSX cross-identified sample (Sect.~\ref{select}) indicates that the
carbon-star contamination is about 23--39\%. Therefore, when selecting
MSX \emph{iiia} sources, using [A]$-$[D] colors (similar to the Messineo sample), 
we find 61\% to be in the oxygen-rich region of the IRAS 2CD (see Table~\ref{t4}).
Assuming that carbon-rich
sources have all their oxygen locked up in carbon-monoxide molecules and thus 
never harbor SiO masers, this suggests that 1) almost all oxygen-rich sources
selected using \emph{iiia} would show detectable 86 GHz SiO maser
emission, and 2) the complicated selection on [A]$-$[D] color
by \citet{messineo02}, which includes flux and variability
information, was likely not much better than simply selecting on
MSX color as they both yield very similar detection rates.
  As individual histogram bins at the blue (or red) end
of the histograms in Fig.~\ref{f11} typically
indicate an 80\% (or 50\%) detection rate, there may be a small trend
of decreasing SiO maser detection rate for redder sources. However, the
range in MSX color is too small, and the uncertainties too large,
to make a definite statement about this.

When compared to the MSX PSC 2.3, \citet{messineo02}'s list of targets
resulted in 204 cross-identifications among the 271 SiO maser
detections and 109 cross-identifications among the 173 targets without
detected SiO maser emission\footnote{Instead of the 313 total sources 
here, \citet{messineo04} found 379 cross-identifications total within
15\arcsec\ radius using MSX PSC version 1.2.}.  The targets without
cross-identifications must be targets that could only be selected
using the more sensitive ISOGAL data -- thus a sample of generally
weaker infrared sources. As relatively more SiO maser detections are
found in the brighter infrared MSX selection this probably confirms
that brighter infrared sources are more likely to show detectable SiO
maser emission \citep[e.g.,][]{messineo02}.

A more careful study can be done with a larger sample, using a larger
color range. Here we have shown how such a sample selected on 
MSX colors (which are directly related to IRAS colors in the \citet{vdveen88}
2CD) can be obtained quite easily compared to previous selections in areas 
where IRAS was confused.

\section{Possible Applications and Future Plans}\label{plans}

It should not be too difficult to find possible applications for this
new tool. Almost any project using IRAS color region selection toward
CSEs in regions on the sky where MSX data is superior to IRAS data can
make use of this new tool. Moreover, particularly in the Galactic
plane, new bulge and center projects can be attempted that were
previously impossible because of the confusion in the IRAS data. For example,
one should be able to find a population of (proto-)planetary nebulae
hidden deeply in the nucleus of the Galaxy and study how their CSEs are
affected by the more hostile local environment. Another example would be
to select thousands of targets to survey for SiO maser emission in the
Galactic plane. Such a survey would e.g.\ yield a map of Galactic rotation based
on a few thousand point masses; gas is much more susceptible to
non-gravitational forces. Stars and CSEs in regions too crowded for
IRAS can be much better classified using MSX colors in terms of IRAS
colors, etc. We leave it to the readers to find further applications in
their own areas of expertise.

A straightforward future plan would be to compare and derive similar
tools using the all-sky AKARI mission survey \citep[formerly
ASTRO-F;][]{murakami07} with respect to the all-sky IRAS survey, but
at much higher angular resolution. The AKARI PSC is expected to become
available sometime after the summer of 2009. We have already commented
on extending this study to the combined GLIMPSE and MIPSGAL data. We
have begun to cross-identify the GLIMPSE and MSX PSCs, but the MIPSGAL
(24 $\mu$m) PSC needed to obtain Spitzer data longward of 8 $\mu$m
has been unfortunately delayed beyond the original release date.

\clearpage\section{Summary}

Using cross-identifications between the IRAS and MSX source catalogs
we have derived color ranges and color regions in MSX 2CDs that
correspond to color regions along the ``evolutionary track'' or the
CSE-sequence in the IRAS 2CD. Oxygen-rich objects in MSX color regions
\emph{i}, \emph{ii}, \emph{iiia}, \emph{iiib}, \emph{iv} and \emph{v}
can be expected to have similar CSEs and CSE properties as oxygen-rich
objects in IRAS color regions I, II, IIIa, IIIb, IV and V,
respectively. This tool should be useful in selecting sources with
different CSEs, especially in regions where selection using IRAS data
is unsatisfactory such as the (IRAS) confused Galactic bulge and plane
regions.

However, it is impossible to distinguish between carbon-rich and
oxygen-rich sources based on MSX colors alone; e.g.\ IRAS or
near-infrared K and/or L band data is needed for that. This means that
when only considering MSX colors, carbon-rich objects and oxygen-rich
objects populate the same MSX regions, in particular \emph{i},
\emph{ii} and \emph{iiia}. Selecting sources with a negative OC-index
will increase the fraction of carbon-rich objects, but also deselect
the low (T $<$ 400 K) temperature carbon-rich objects. The reverse 
selection is  not likely: that one increases the fraction of oxygen-rich
objects in a sample by selecting sources with positive OC-index, unless 
only cool sources are considered.

We also seem to have revealed an error in the photometry of the MSX E 
filter band. The flux density seems a factor $\sim$ 2 too high, 
or $\sim$ 0.5--1 magnitude too bright in [E], for weak A filter band 
sources in the MSX point source catalog version 2.3. This error does 
not seem to be evident in MSX PSC version 1.2, which has a smaller number 
of weak sources. The reason for this discrepancy is
  probably due to the band E detector and a different handling of the 
photometry.

\vfill

\section*{Acknowledgments} 
We thank 
Maria Messineo and Roberto Ortiz for providing data, insight and
healthy discussions. SMC acknowledges support from the ``Research
Experiences for Undergraduates'' program, which is funded by the
National Science Foundation. The National Radio Astronomy Observatory
is a facility of the National Science Foundation operated under
cooperative agreement by Associated Universities, Inc.

{\it Facilities:} \facility{IRAS ()}, \facility{MSX ()}

\end{document}